\documentclass[a4paper,aps,preprintnumbers,twocolumn,amsmath,amssymb,pra,showpacs]{revtex4}
\usepackage{bbm}
\usepackage{mathrsfs}
\usepackage{amsmath}
\usepackage[dvips]{graphicx}
\usepackage{epsfig}
\usepackage[dvips]{graphics,graphicx}

\newcommand{\be}{\begin{equation}}
\newcommand{\ee}{\end{equation}}
\newcommand{\bea}{\begin{eqnarray}}
\newcommand{\eea}{\end{eqnarray}}
\newcommand{\bes}{\begin{split}}
\newcommand{\ees}{\end{split}}

\newcommand{\tr}{\operatorname{Tr}}

\usepackage[dvips]{graphics,graphicx}

\usepackage{color}

\begin{document}
\title{Thermalization  in Systems with Bipartite  Eigenmode Entanglement}
\author{Ming-Chiang Chung}
\affiliation{Physics Division, National Center for Theoretical Science, Hsinchu, 30013, Taiwan}
\affiliation{Institute of Physics, Academia Sinica, Taipei 11529, Taiwan}
\author{A. Iucci}
\affiliation{Instituto de F\'isica de La Plata (IFLP) - CONICET
  and Departamento de F\'isica, Universidad Nacional de La Plata, cc
  67, 1900 La Plata, Argentina}
\author{M. A. Cazalilla}
\affiliation{Centro de F\'isica de Materiales CSIC-UPV/EHU}
\affiliation{Donostia International Physics Center (DIPC),
 Paseo Manuel de Lardizabal 5, E-20018 San Sebastian, Spain}

\begin{abstract}
It is analytically shown that the asymptotic correlations  in exactly solvable models
following a quantum quench can behave essentially as thermal correlations
provided the entanglement between two eigenmodes is sufficiently strong.
We provide one example and one counter example of this observation. The
example illustrates the fact that the  thermal correlations arise
from initial states where the entanglement between the eigenmodes stems from the existence of a
large energy gap in the initial state. On the other hand, the counter-example shows that when the bi-partite entanglement of the eigenmodes stems from interactions that do not open a gap,  the correlations at asymptotically long times are non-thermal.
We also show that the thermal behavior concerns only the asymptotic correlation functions, as the difference with an
actual thermal ensemble can be observed measuring the energy fluctuations of  the system.
The latter observation implies a breakdown of the fluctuation-dissipation theorem.
\end{abstract}


\date{\today}
\maketitle

\section{Introduction}\label{sec:intro}

 Experiments with ultracold gases~\cite{Weiss,Schmiedmayer,Bloch,Greiner,CazalillaRigol}, and, in particular,
 the ongoing efforts to build quantum emulators using ultracold atoms loaded in optical lattices,
 have aroused much interest in understanding the thermalization mechanisms of integrable
 models\cite{Weiss,Rigol07,Rigol06,Rigol11,Sengupta,Cazalilla,attention,MiguelAnibalMing,
Polkovnikovetal}.
 The latter can be used as simple systems to validate a quantum  emulator by comparing the
 outcome of the experiment to the exact solution, prior to using the emulator to study other,
 more complex models, for which no exact solutions are known.

 However, in order to understand the outcome of a simulation of an integrable or exactly solvable
 model it is important to understand the effect of  the initial conditions.  Since quantum emulators of ultracold
 atoms are largely isolated systems and its evolution is essentially unitary, it becomes necessary to understand the conditions
 under which the asymptotic state of the system can be described by a standard statistical (i.e. Gibbsian) ensemble,
 or, as it was pointed out recently by Rigol and coworkers~\cite{Rigol07}, must be described by a generalized Gibbs
 ensemble (GGE). The latter captures the fact that the existence of a non-trivial set of integrals of motion  strongly constraint the
 non-equilibrium dynamics of the system.

  In Ref.~\cite{MiguelAnibalMing}, we showed that the GGE can be
 analytically derived for exactly solvable models and a general class of
 initial states. In particular, we showed that dephasing between different modes makes equal time
 correlation functions of both local and non-local operators non-ergodic, in the sense that in the
 thermodynamic limit, their infinite time limit  only depends on the occupation of the eigenmodes
 in the initial state only. Thus, the asymptotic values
 of those correlation functions can be equivalently obtained by assuming that the correlations with
 other eigenmodes  produce an effective temperature. This yields a description of the asymptotic
 correlations that is entirely equivalent to the GGE.

Nevertheless, it was also noticed in Refs.~\cite{Rigol06,Rigol11}
that certain kinds of initial states can lead to asymptotic values of the observables that are essentially indistinguishable from those computed with a standard thermal Gibbs ensemble. Other cases of (pre-) thermalization have been 
found in integrable~\cite{grisins} and non integrable~\cite{Rigol_ETH,afho10,gring} systems for particular classes of initial conditions.
Recently, Mitra and Giamarchi~\cite{mitraGiam} also showed that the adiabatic
introduction of  a non-linearity following a quantum quench in the Luttinger model,
can lead to thermalization as described by the standard Gibbs ensemble.
The authors of Ref.~\cite{mitraGiam} emphasized the
importance of  ``mode coupling'' for thermalization.
In this work, we show that for certain classes of quenches for which two sets of
modes are strongly entangled in the initial state, the GGE ensemble can
be arbitrarily close to the standard Gibbs ensemble.
Using  the methods of Ref.~\cite{MiguelAnibalMing}, we find a simple instance of the mechanism by which
initial states can lead to correlations that essentially look thermal. As we show below, this happens
when the eigenmodes of the Hamiltonian that describes the evolution of a system following a quantum quench
have a certain kind of bi-partite quantum correlations (i.e. entanglement) in the initial state. As an application,
we find an analytical explanation for the numerical observations first reported in Ref.~\cite{Rigol06}.
This fact could be used as a simple protocol to produce asymptotic
correlations that \emph{essentially} look as (rather high temperature) thermal
correlations  in exactly solvable systems.  Furthermore,  we will show below that 
the effective  temperature that characterizes the asymptotic thermal correlations can be  related to the
entanglement spectrum of a subset of entangled modes, which  means that the latter is accessible experimentally 
by measuring the effective temperature that characterizes the correlations at long times after the quench.  

 Before illustrating the above point with one example and one counter example,
let us first describe the general set up which will be  addressed below.
Consider  a system containing two  subsystems $A$ and $B$  that are initially coupled
together. For times $t \leq 0$, the system is described by  a Hamiltonian of the form $H_0 = H_A + H_B
+ H_{AB}$ where $H_A$,$H_B$ and $H_{AB}$ are quadratic in some eigenmodes $\{a_k, b_k\}$
which carry a  quantum number $k$ (which forms an continuum in the thermodynamic limit)
and can be fermionic  or bosonic, i.e.
\begin{align}
H_A &= \sum_k \varepsilon_A(k) a_k^{\dagger} a_k,\\
H_B &=  \sum_{k} \varepsilon_B(k) b_k^{\dagger} b_k,\\
H_{AB} &= \sum_{k} \Delta_{AB}(k) \left[ a^{\dag}_k b_{k} + b^{\dag}_k a_k \right].
\end{align}
The dispersion relations are assumed such that $\epsilon_A(k) \neq \epsilon_B(k)$ for essentially all $k$,
which is required (see below) for dephasing between the two subsystems to occur as $t\to +\infty$.
We can assume that the system is prepared in an initial state in contact with a thermal reservoir at
a temperature $T$,  i.e. $\rho_0 = Z^{-1}_0 \: e^{-H_0/T}$ (such that $\mathrm{Tr} \, \rho_0 = 1$).
For $t > 0$, the coupling between the two
subsystems $H_{AB}$ disappears, and the two subsystems evolve unitarity
and uncoupled, according to the Hamiltonian:
\be
H=  H_A + H_B.
\ee
The existence of the coupling $H_{AB}$ for all $t \leq 0$ implies that in the initial
state, $\rho_0$, there are  correlations (i.e. bi-partite entanglement)
between the eigenmodes, i.e. $\langle a^{\dag}_k b_k \rangle \neq 0$.

 According to  the conjecture of Rigol \emph{et al.}~\cite{Rigol07}, the asymptotic state of the system 
 can be described by a `generalized' Gibbs ensemble (GGE) density matrix that is obtained 
 using the maximum entropy principle taking into account that  the system dynamics  is constrained by the
existence of  the set of  integrals of motion given by $I_a(k) = a^{\dag}(k) a(k)$ and $I_b(k) = b^{\dag}(k) b(k)$. 
The GGE density matrix thus obtained reads:
\be
\rho_{\mathrm{GGE}} = Z^{-1}_{\mathrm{GGE}}   \exp\left\{ - \sum_{k}  \left[ \alpha(k) a_k^{\dagger} a_k +   \beta(k)
  b_k^{\dagger} b_k \right]\right\} , \label{eq:GGE}
\ee
where  the Lagrange multipliers are determined by the initial conditions, i.e.
$\alpha(k) = \ln [(1 \pm n^a(k))/n^a(k)]$ and  $\alpha(k) = \ln
[(1 \pm n^b(k))/n^b(k)]$, with $n^a(k)$ and $n^b(k)$ given by~\eqref{eq:na} and
\eqref{eq:nb} (the $+$ applies to bosonic and the $-$ to fermonic modes).

 Alternatively, one can arrive at an equivalent result by a different route~\cite{MiguelAnibalMing}.
 Let us first consider the expansion of a local operator in terms of the eigenmodes of $H$:
\be
   O(x,t) = \sum_{k } \left[   \phi^A_k(x) e^{-i \varepsilon_A(k) t} a_k +
  \phi^B_k(x)  e^{-i \varepsilon_B(k) t} b_k \right].
\ee
At asymptotically long times after the quantum quench, provided $\epsilon_A(k)\neq \epsilon_B(k)$
and certain conditions of smoothness are met, dephasing renders the two-point correlation
function $\langle O^{\dag}(x,t) O(0,t) \rangle$  to the following form~\cite{MiguelAnibalMing}:
\be
  \begin{split}
  \lim_{t \rightarrow \infty}  & \langle O^{\dagger}(x,t) O(0,t) \rangle
      =  \sum_{k}   \left[ \phi^A_k (x)  \right]^{\ast} \phi^A_k(0) \langle
    a_k^{\dagger} a_k \rangle  \\  & \qquad +  \sum_{k} \left[ \phi^B_k(x)\right]^{\ast} \phi^B_k(0) \langle
    b_k^{\dagger} b_k \rangle, \label{eq:sumk}
  \end{split}
\ee
Thus, we see  that the asymptotic correlations of $O(x)$ depend only
on the eigenmode occupation in the initial state, a behavior that
has been termed non-ergodic in Ref.~\cite{MiguelAnibalMing}. 
The above sum over $k$ in Eq.~\eqref{eq:sumk} allows to define
a mode-dependent temperature for each mode~\cite{MiguelAnibalMing}.
Indeed, this statement is equivalent to the GGE (cf. Eq.~\ref{eq:GGE})
for a broad class of (Gaussian) initial states (see Ref.~\cite{MiguelAnibalMing} and below).
Thus, it follows that:
\begin{equation}
\lim_{t\to +\infty} \langle O^{\dag}(x,t) O(0,t) \rangle = \langle O^{\dag}(x) O(0) \rangle_{\mathrm{GGE}}.
\end{equation}
The above result, valid for local operators,  can be combined  with Wick's theorem to show that the
asymptotic behavior of non-local operators is also described by the GGE \cite{MiguelAnibalMing}.

 Alternatively,  when the correlations are
bi-partite, we can regard the effective temperature for the modes in the subsystem $A$ as due
to their entanglement with the modes in the subsystem $B$ (and viceversa).
Thus, whenever we are dealing with $\langle
a^{\dag}_k a_k\rangle =  \tr \rho_0 a_k^{\dagger} a_k$ or
$\langle b^{\dag}_k b_k \rangle=\tr \rho_0 b_k^{\dagger} b_k$, we can trace out one
of the subsystems, and write:
\bea
    n^a(k) &= \langle  a_k^{\dagger} a_k \rangle  =
    \tr \rho_{A}  a_k^{\dagger} a_k =  \tr \rho_{GGE}  \: a_k^{\dagger} a_k ,  \label{eq:na} \\
    n^{b}(k) &= \langle b_k^{\dagger} b_k \rangle   = \tr \rho_{B}  b_k^{\dagger} b_k =
    \tr \rho_{GGE}  \: b_k^{\dagger} b_k, \label{eq:nb}
\eea
where $\rho_A = \tr_{B} \rho_0$ and $\rho_B = \tr_{A} \rho_0$. Therefore,
the GGE density matrix can be written as:
\be
  \rho_{\mathrm{GGE}} = \rho_A \otimes \rho_B. \label{eq:corresp}
 \ee
We can regard the result in Eq.~\eqref{eq:corresp} as a way to relate the density matrix of the
 GGE  ensemble to the reduced density matrices of the subsystems $A$ and $B$.  Furthermore, since both
 $\rho_A$ and $\rho_B$ are hermitian,  it is possible to write these objects as follows~\cite{Peschel,Li/Haldane08}:
\begin{equation}
\rho_{A(B)} = \frac{e^{-\mathcal{H}_{A(B)}} }{Z_{A(B)}},
\end{equation}
where we have introduced the entanglement Hamiltonian of the  $A$ ($B$) subsystem $\mathcal{H}_{A(B)}$,
which is also a hermitian operator. Thus, we see $\rho_{\mathrm{GGE}}$ is determined by
the total entanglement Hamiltonian, $\mathcal{H} = \mathcal{H}_A + \mathcal{H}_B$.

The reduced density matrix  describing a
subsystem $A$ of either a pure state or a thermal mixed state  plays
an important role in quantum information theory applied to condensed matter systems~\cite{Peschel}.
For a pure state,   the von Neumann entropy  $S_{A} = -\tr \rho_{A}
\log_2 \rho_{A}$ measures the entanglement between  two subsystems $A$
and $B$. The latter  can be expressed in terms of the entaglement spectrum of $\mathcal{H}_A$.
Recently, the von Neuman entropy and entanglement spectrum have become an important tool, 
as they can be used to characterize topological  quantum
phases in various kinds of quantum systems, such like graphene~\cite{ChungJhuChenYip}, topological
insulators \cite{QiLudwig}, and quantum spin chains ~\cite{Pollmann}. In this context, an
important question that has been  addressed in recent times are the conditions under the
entanglement Hamiltonian  $\mathcal{H}_A$ can be proportional to the subsystem Hamiltonian $H_A$.
Some examples of this fact
been discussed in the literature~\cite{QiLudwig,Lauchli10, Schliemann11, Poilblanc10,PeschelChung11}.
As we show in the example below (see section~\ref{sec:example1}), when this happens to be the case in 
a system like the one described above, we can expect the asymptotic correlations  
after the quantum quench to become essentially thermal. 

  Using the methods of Refs.~\cite{ReducedDensity}, the entanglement Hamiltonian $\mathcal{H}_{A(B)}$
can be determined for a (Gaussian) initial state of the form $\rho_0 = Z^{-1}_0\: e^{-H_0/T}$
($\rho_0 = |\Phi_0\rangle \langle  \Phi_0 | /\langle \Phi_0 | \Phi_0 \rangle$, where $|\Phi_0\rangle$
is the ground state of $H_0$ at $T = 0$). Thus~\cite{ReducedDensity},
\begin{align}
 \mathcal{H}_A &= \sum_k \ln \left[(1 \pm n^a(k))/n^a(k)\right] a_k^{\dagger} a_k, \label{rhoA} \\
\mathcal {H}_B &=   \sum_k \ln \left[(1 \pm n^b(k))/n^b(k)\right] b_k^{\dagger} b_k,
\end{align}
which, by comparison with Eq.~\eqref{eq:GGE}, allows us to identify the Lagrange multipliers
$\alpha(k)$ and $\beta(k)$ of the GGE with the entanglement spectrum of the
subsystems $A$ and $B$.

 Thus, the entanglement spectrum determines the asymptotic state following a quantum quench.
Similar ideas have been discussed  by Qi {\it et. al.} for the particular case of two-coupled edge states
using boundary conformal field theory~\cite{QiLudwig}. Conversely,  provided
Lagrange multipliers $\alpha(k)$ and $\beta(k)$ could determined experimentally,
we would be able to access the entanglement spectrum and the von Neumann
entropy of the subsystems $A$ and $B$. However, in actual experiments it may be difficult to obtain the full
functional dependence of $\alpha(k)$ and $\beta(k)$. Thus, below we shall focus on two
cases where the entanglement spectrum takes a simple form, which may be easier to
measure experimentally.

 The rest of this work is organized as follows. In the following section, using the above results,
we provide an example of the case in which the asymptotic correlations are essentially thermal,
which we show to be a consequence of the entanglement hamiltonians $\mathcal{H}_{A(B)}$ to be proportional to 
the subsystem Hamiltonian, $H_{A(B)}$. In section~\ref{sec:example2}, we provide a counter-example of the fact that 
bi-partite entanglement does not always lead to thermal correlations. This counter-example illustrates the observation that thermal correlations appear provided the bi-partite entanglement arises due from a gap in the spectrum of the Hamiltonian
that determines the initial state. Finally, in section~\ref{sec:conclusion}, we provide a discussion of our results
and show that, even when the correlations look essentially thermal, there are certain observables like the energy fluctuations
that still differ from  their thermal values, a fact that signals a breakdown of the fluctuation-dissipation theorem in the asymptotic
state. The appendix contains some technical details regarding a continuum version of the model discussed 
section~\ref{sec:example1}.

\section{Example}\label{sec:example1}

 Let us first consider a model that has been  numerically  studied earlier by Rigol and coworkers \cite{Rigol06}. The model describes  a system of hard-core bosons in 1D that initially (i.e. for $t \leq 0$) move in the presence of superlattice potential. The hard-core bosons in 1D can be  treated using a Jordan Wigner transformation~\cite{giamBook,lieb_antiferromagnetic_chain}, which maps the hard-core bosons to non-interacting fermions and, in the case of a superlattice of strength $\Delta$,  leads to the following quadratic Hamiltonian:
\be \label{HRigol}
  H_0 = - \sum_{j=1}^{L} \left( f_j^{\dagger} f_{j+1} + f^{\dag}_{j+1} f_j \right) + \Delta \sum_{j=1}^{L}  (-1)^j    f_j^{\dagger} f_j,
\ee
where $f_{j}^{\dagger}$ and $f_j$ are creation and annihilation operators for spinless fermions at site $j$ ($j=1,\ldots,L$, 
for a lattice of $L$ sites).  Rigol \emph{et al.}~\cite{Rigol06} numerically found that, starting from the ground state of $H_0$, if  the superlattice term $\propto \Delta$ is suddenly switched off at $t = 0$, and the system is allowed to evolve unitarity according to
\be
   H = -  \sum_{j=1}^{L} \left( f_j^{\dagger} f_{j+1} + f^{\dag}_{j+1} f_j \right), \label{eq:postham}
\ee
the long-time behavior of the momentum distribution can described by a standard Gibbs canonical ensemble,
\be
 \rho = \frac{1}{Z} e^{-H/T_\mathrm{eff}}, \label{eq:thermal}
\ee
for which the effective temperature, $T_\mathrm{eff}$,  was found to approach $\Delta/2$ for $\Delta\gtrsim 1$. In what follows,
we will analytically demonstrate that this numerical observation indeed  follows from the existence of a strong
bi-partite entanglement between two sets of  eigenmodes of $H$.

 We  begin by  Fourier transforming $H_0$ and $H$ by using~\footnote{We use periodic
boundary conditions here whereas Rigol and coworkers used  open boundary conditions~\cite{Rigol06}. The difference is irrelevant
in the thermodynamic limit, where the methods of Ref.~\cite{MiguelAnibalMing} apply.}
\be
f_k = \frac{1}{\sqrt{L}}  \sum_{j=1}^L  e^{-i  k x_i} f_i,
\ee
with $x_j  = j$,  the Hamiltonian (\ref{HRigol}) can be written as
\bea
   H_0 & =&  H + \Delta\sum_{k}\left( f_k^{\dagger} f_{k+\pi}
           + f^{\dagger}_{k+\pi} f_{k}  \right), \label{eq:HRigolk}\\
  H & = & \sum_{k} \omega_k \left( f_k^{\dagger} f_{k}-
   f_{k+\pi}^{\dagger} f_{k+\pi} \right) \label{H}
\eea
where $\omega_k = - 2 \cos{k}$ and  $-\pi/2 < k \leq \pi/2$. The Hamiltonian describing the state the system at $t < 0$,
namely $H_0$,  can be brought to diagonal form by means of the following canonical transformation:
\begin{align}
\gamma_k&= \cos\theta_k f_k + \sin\theta_k f_{k+\pi}, \label{eq:bogol1}\\
\delta_k  &= -\sin{\theta_k} f_k + \cos{\theta_k} f_{k+\pi}, \label{eq:bogol2}
\end{align}
with $\tan 2\theta_k = \frac{\Delta}{\omega_k}$. Hence,
\be
H_0= \sum_{k} E_k \left(\gamma_k^{\dagger}
 \gamma_k - \delta_k^{\dagger} \delta_k\right), \label{HGD}
\ee
where   $E_k=\sqrt{\omega^2_k + \Delta^2}$. Note that the  transformation
in \eqref{eq:bogol1} and \eqref{eq:bogol2} implies the existence of strong bi-partite
quantum correlations (i.e. entanglement) between the modes at $k$ and $k +\pi$, which manifest in, e.g.
$\langle f^{\dag}_k f_{k+\pi}\rangle = -\frac{1}{2} \sin 2 \theta_k  = - \frac{\Delta}{2 E_k} \neq 0$.

 As discussed in Refs.~\cite{Sengupta,MiguelAnibalMing},  the asymptotic momentum distribution of the
hardcore bosons for $t \to +\infty$  can be obtained from the Fourier transform of  the one-particle correlation function of the bosons,
which in turn can be written as a Toeplitz determinant involving correlation two-point correlations
of the Jordan-Wigner fermions:
\begin{align}
\lim_{t\to+\infty} g^{(1)}(x_i-x_j,t) &=&  \left|
\begin{array}{cccc}
a_0 & a_1 & \cdots & a_{-n+1}\\
a_1 &  a_0 & \cdots & a_{-n+2}\\
\vdots & \vdots & \ddots & \vdots \\
a_{n-1} & a_{n-2} & \cdots & a_0
\end{array}
\right|,\label{eq:toeplitz}
\end{align}
where
\be
a_{i-j+1} = - \lim_{t\to+\infty}\langle A_i(t) B_j(t)\rangle.
\ee
 The thermodynamic limit is implicitly understood in the above expressions; 
 we have also introduced the notations: $A_j = f^{\dag}_j + f_j$ and $B_j = f^{\dag}_j - f_j$.
The above correlation function of the (local) operators $A_i$ and $B_j$ can be shown
to be:
\begin{align}
a_{i-j+1} &= \frac{1}{L} \sum_k e^{-i k (x_i-x_j)} \left[ 2 n(k) -1 \right] \nonumber\\
 &+ \frac{(-1)^{i-j}}{L} \sum_k e^{-i k (x_i-x_j)} \left[ 2 n(k+\pi) -1 \right]  
\end{align}
Next, we use that ($K = 0,\pi$):
\begin{align}
n(k+K)  = \langle f^{\dag}_{k+K} f_{k+K} \rangle &=  \mathrm{Tr}\:  \rho_0 f^{\dag}_{k+K} f_{k+K} \\
&=  \mathrm{Tr} \:  \rho_K \:
f^{\dag}_{k+K} f_{k+K},
\end{align}
where $\rho_0 = |\Phi_0 \rangle \langle \Phi_0 |$,  $\Phi_0$ being the initial state, that is, the ground state of $H_0$ (cf.
Eq.\eqref{HRigol}) and the reduced density matrices:
\be
\rho_{K}  =
 \mathrm{Tr}_{k \in S_K} \rho_0,
\ee
where $S_{0} = \left( -\pi/2, +\pi/2 \right]$, and $S_{\pi}  = \left(-\pi,+\pi \right] - S_{0}$.
In other words, the asymptotic correlations can obtained from the reduced density matrix
resulting from tracing one one the two sets of entangled modes with $k$ belonging to either
$S_{0}$ or $S_{\pi}$.

 As explained in section~\ref{sec:intro}, the reduced matrices $\rho_{K=0,\pi}$ can be
obtained analytically in terms of the occupation numbers of $n(k+K) = \langle
f^{\dag}_{k+K} f_{k+K} \rangle$ in the initial state.  Using  \eqref{eq:bogol1} and \eqref{eq:bogol2},
we find:
\begin{align}
n(k)&=\frac{1}{2}\left( 1 - \frac{\omega_k}{E_k}\right),\\
n(k+\pi)&=\frac{1}{2}\left( 1 + \frac{\omega_k}{E_k}\right),
\end{align}
and, following the discussion in  section~\ref{sec:intro},  the Lagrange 
multipliers determining the GGE density matrix
read:
\be
  \begin{split}
   \alpha(k)  = & \ln\left[(1 - n(k))/n(k)\right] \\
    =&  \ln\left[\frac{E_k+\omega_k}{E_k-\omega_k} \right]  \\
   \beta(k)  = & \ln\left[(1 - n(k+\pi))/n(k+\pi)\right]  \\ = &
   \ln\left[\frac{E_k-\omega_k}{E_k+\omega_k} \right].
  \end{split} \label{alphabeta}
 \ee
 For $\Delta \gg \omega_k$, $E_k$ can be approximated  by
$\Delta$, and therefore $\alpha(k) \simeq 2\omega_k/\Delta \simeq - \beta(k)$.
Thus, the GGE density matrix, Eq.~\eqref{eq:GGE} reduces to a standard Gibbs
ensemble, Eq.~\eqref{eq:thermal}  with
\be
T_\mathrm{eff} \simeq \Delta/2,\label{eq:efft}
\ee
which is in agreement with the numerical observations of Rigol and coworkers~\cite{Rigol06}.
However, it is important to note that the above thermal ensemble and the result of  Eq.~\eqref{eq:efft}
is only an approximation to the actual GGE ensemble determined by the Lagrange multipliers in Eq.~\eqref{alphabeta}. However,
this approximation becomes better and better for larger values of $\Delta$, which implies that numerically
(and experimentally) the GGE and a standard thermal Gibbs ensemble will be essentially indistinguishable.

 It is worth noting that the above results are also relevant for a special limit an integrable field theory in one dimension,
namely the sine-Gordon model:
\begin{align}
H_\text{sG} &= H_0 -  \frac{v g}{\pi a^2_0} \int dx \,
\cos 2 \phi, \label{eq:sG}\\
H_0 &= \frac{v}{2\pi} \int  dx \, \left[ K^{-1}\left( \partial_x
\phi\right)^2 + K \left(\partial_x \theta \right)^2 \right],
\end{align}
where $a_0$ is a short-distance cut-off, and the phase and density fields $\theta(x)$ and
$\phi(x)$, are canonically conjugated in the sense that the obey:
$[\phi(x), \partial_{x'} \theta(x')] = i \pi \delta(x-x')$; $v$ is the speed of
sound and $K$ is a dimensionless parameter that  determines the ground correlations
of the system. Indeed, in equilibrium, the model
exhibits two phases for $g>0$, namely, a gapped phase for $K<2$ and gapless
phase for $K\geq 2$~\cite{giamBook}. For $K  = 1$, as described in the the Appendix,
this model can be mapped onto a model of massive fermions in 1D:
\begin{multline}
H_{sG}  =\sum_{p}\varepsilon_{p}^{0}\left[\psi_{R}^{\dagger}(p)\psi_{R}(p)-\psi_{L}^{\dagger}(p)\psi_{L}(p)\right] \\
+\Delta\sum_{p}\left[\psi_{R}^{\dagger}(p)\psi_{L}(p)+\psi_{L}^{\dagger}(p)\psi_{R}(p)\right],\label{eq:sG_fermions}
\end{multline}
with  $\varepsilon_{p}^{0}= vp$ and $\Delta \propto g$.  The model in Eq.~\eqref{eq:sG_fermions} can be regarded
as the continuum limit of the model of Eq.~\eqref{eq:HRigolk}.

For large values of the gap, $\Delta\gg v a_0^{-1}$, the multipliers $\alpha(p)$ and $\beta(p)$ (see calculation in the
Appendix)  become proportional to the dispersion relation of the Hamiltonian that governs the time
evolution, $\pm\varepsilon^0_p$ respectively, with an effective temperature $T_\mathrm{eff}$ (it is the same for the two branches of fermions) given by
\begin{equation}
T_{\mathrm{eff}}=\frac{\Delta}{2 \tanh\frac{\Delta}{2 T}}
\end{equation}
We notice that  the effective temperature  depends on the temperature
of the initial thermal state. This effective ``final'' temperature is
always higher than the initial temperature because it contains the
bi-partite entanglement between $k$ modes. For high initial temperatures the effective temperature to which the system thermalizes is the same as the initial one. In the case of a zero-temperature initial state, the effective temperature results $T_\mathrm{eff}=\Delta/2$ similarly to the case of $XY$ model studied previously.

Generally speaking, the analysis described above is also related to the discussion of the
conditions under which the entanglement Hamiltonian $\mathcal{H}_{A(B)}$   and the Hamiltonian of the subsystem
$H_{A(B)}$ are (approximately) proportional to each other. If this is the case, then, according to the discussion of section~\ref{sec:intro},
the GGE density matrix, Eq.~\ref{eq:GGE}, will be well approximated by the thermal density matrix of~\eqref{eq:thermal}.
Indeed, recently Peschel and Chung~\cite{PeschelChung11}   addressed the problem of the proportionality
 between $\mathcal{H}_{A(B)}$ and $H_{A(B)}$. By considering a model of two
 two species of fermions with opposite dispersion $\omega_k$ and coupling $\Delta$, which
leads to an energy spectrum for the coupled system with a gap of magnitude $2\Delta$.  Using perturbation theory,
they showed  for $\Delta \gg \omega_k$ that
${\cal H}_A \simeq 2/\Delta \sum_k  \omega_k a_k^{\dagger} a_k = (2/\Delta) H_A$
and ${\cal H}_B \simeq 2/\Delta \sum_k  (-\omega_k) b_k^{\dagger} b_k = (2/\Delta)
H_B$.  The thermal correlations obtained here can be thus regarded as direct consequence
of this result when we apply it to a quantum quench where the coupling $\Delta$ is switched off at $t=0$ and
we exploit the relationship between the GGE and the reduced density matrices $\rho_{A(B)}$
described in section~\ref{sec:intro}.

Another interesting consequence of the above result is the possibility to use quantum quenches to
prepare systems with exactly solvable dynamics in states whose correlations will become
indistinguishable from thermal correlations after the quench. However, it is unfortunate
that the requirement of a large gap (i.e. the condition that $\Delta\gg 1$), implies that thermal states
that can be thus obtained are characterized by extremely high effective temperatures (cf. Eq.~\ref{eq:efft}).

\section{Counter-example}\label{sec:example2}

The above result on the emergence of thermal behavior at long times stems from the existence of strong
bi-partite entanglement in the initial state. However, as we show in this section, the existence of such entanglement
is not a sufficient condition for the emergence of thermal correlations. Indeed, whenever the initial state  is gapless,
no thermal behavior can be expected even for large entanglement. The Luttinger model~\cite{giamBook}
is one such example, as we show below.

Let us consider a quantum quench in the Luttinger model (LM)
\cite{Cazalilla}. The initial state is assume to be a mixed thermal state $\rho_0 = Z^{-1}_0
e^{-H^{LM}_0/T}$. The initial Hamiltonian $H_0^{LM} = H + H_{\mathrm{int}}$, where
\bea \label{HLM}
   H^{LM} & = & \frac{2\pi v_F}{L} \sum_{k>0}  \rho_R(k)
   \rho_R(-k) + \rho_L(-k) \rho_L(k), \quad\quad\\
   H_{\mathrm{int}}  & = &  \frac{ \Delta}{L}\sum_{k>0} \rho_R(k) \rho_L(-k) +
   \rho_R(-k) \rho_L(k).
\eea
In the above expression,  $\rho_{R(L)}$ is the density of the right (left) moving
fermions in the Luttinger model, which  propagate with  Fermi velocity ($v_F$)~\cite{Cazalilla}.  The densities obey the
commutation rule $[\rho_{\alpha}(k),\rho_ {\beta}(k')] = kL/2\pi
\delta_{k+k'} \delta_{\alpha, \beta}  (\alpha, \beta = R,L)$.
Therefore we can define two pairs of bosonic operators: $\rho_L(k) =
\sqrt{kL/2\pi} a_k^{\dagger}$, $\rho_R(k) =
\sqrt{kL/2\pi} b_k^{\dagger}$ and $\rho_L(-k) = \rho_L^{\dagger}(k)$,
$\rho_R(-k) = \rho_R^{\dagger}(k)$. Thus, the  Hamiltonian describing the system at $t < 0$ 
can be written as
\be \label{HLMab}
    H_0^{LM} = \sum_{k>0} \left[  v_F k (a_k^{\dagger} a_k + b_k^{\dagger} b_k) +
    \frac{\Delta}{2\pi} k (a_k^{\dagger}  b_k^{\dagger} + a_k b_k)\right].
\ee
Note that the LM is different from the  example that has been discussed
in section~\ref{sec:example1} because the term that couples the two subsystems is
proportional to  $k$, while in the previous example (cf. Eq.~\ref{HRigol}) it was  a constant.
This makes the Hamiltonian $H^{LM}_0$ gapless, as we show 
in the following paragraph.

A standard way to diagonalize (\ref{HLMab}) is to introduce a bosonic
canonical transformation:
 $A_k =
\cosh{\phi_k} a_k - \sinh{\phi_k} b_k^{\dagger}$ and $\delta_k =
-\sinh{\phi_k} a_k + \cosh{\phi_k} b_k^{\dagger}$. Choosing $\tanh{
(2\phi_k)}  = - \Delta/2\pi $. The initial Hamiltonian reads now
\be
    H_0^{LM}=  \sum_{k>0} \Omega_k (A_k^{\dagger} A_k + B_k^{\dagger} B_k),
\ee
where $\Omega_k = v_F k \sqrt{(1-(\Delta/2\pi v_F)^2)}$.
Thus, the energy spectrum  of $H^{LM}_0$  is 
is gapless.

According to the discussion of section~\ref{sec:intro} and using the methods of Ref.~\cite{MiguelAnibalMing},
after turning off the  interaction described by $H_{\mathrm{int}} \propto \Delta$ at $t = 0$, the asymptotic
behavior of the correlations can be described by a GGE matrix, which can be written as $\rho_{GGE}$ as
in Eq.~\eqref{eq:GGE} with $\alpha(k)$ and $\beta(k)$ given by the entanglement spectrum of the subsystems
$A$ and $B$,  of modes $a_k, a^{\dag}_k$ and $b_k, b^{\dag}_k$ respectively.

The occupation numbers are: $n^a(k) = n^b(k) = \sinh^2(\phi_k) = 1/2(v_F
k/\Omega_k-1)$.  Hence, $\alpha(k)$  and $\beta(k)$ in Eq.~(\ref{eq:GGE})
are:
\begin{align}
    \alpha(k) &= \ln{\left(\frac{v_F k + \Omega_k}{v_F k
          -\Omega_k}\right)} = \varepsilon,\\
      \beta(k)    &= \ln{\left(\frac{v_F k + \Omega_k}{v_F k
          -\Omega_k}\right)} = \varepsilon, 
\end{align}   
where $\varepsilon = 2[\ln{(2\pi v_F/\Delta + \sqrt{(2\pi v_F/\Delta)^2-1})} ] $ is a constant. 
Hence, the entanglement Hamiltonian take the form:
$\mathcal{H}_A =   \sum_k  \varepsilon \: a^{\dagger} a_k$
and $\mathcal{H}_B =   \sum_k  \varepsilon \: b^{\dagger} b_k$.
Thus, we find that  $\mathcal{H}_{A(B)}$ is not proportional to  $H_{A(B)}$. 
It then follows that  he density matrix of GGE,  $\rho_{GGE} = Z^{-1}_{\mathrm{GGE}}\:
e^{- (\mathcal{H}_A + \mathcal{H}_B)}$  does no longer reduce to a thermal ensemble. 
Thus, we conclude that the existence of bi-partite entanglement in the initial state
is not a sufficient condition for the emergence of asymptotic thermal correlations. An additional condition, 
such us the existence of a gap, is appears to be required.

\section{discussion and conclusion}\label{sec:conclusion}

In this section, we  would like to discuss  a number of points concerning the above results.
The first point is concerns the opposite situation to the one discussed in  section~\ref{sec:example1}.
We could ask what happens if initially the two subsystems $A$ and $B$ are not coupled and, at $t=0$,  
they suddenly become coupled so that entanglement is created.  Can we still expect the asymptotic
correlations following such a quench to be described by an essentially thermal ensemble 
in a certain parameter regime?   The answer is no, as we explain below.

  To address the above question, let us  imagine that, initially, the bosons
are free to hop everywhere and there is no superlattice. However, 
at $t=0$ a superlattice is imposed, say by the sudden application of an extra pair of  counter-propagating 
laser beams. Mathematically, the system at $t \leq 0$ is  described by the Hamiltonian $H$ (cf. Eq.~\ref{H})
and its subsequent evolution at $t > 0$ is described by $H_0$ (cf. Eq.~\ref{eq:HRigolk}). The density matrix of GGE is
determined by the occupation numbers  $n^{\gamma}(k) = \tr \rho_0
\gamma_k^{\dagger} \gamma_k = \sin^2{\theta}_k$  and $n^{\delta}(k) = \tr \rho_0
\delta_k^{\dagger} \delta_k = \cos^2{\theta}_k$, for a half-filled lattice.
 Proceeding as above, the GGE density matrix describing the asymptotic
 correlations is:
\be \label{DMCouple}
   \rho_{GGE} \simeq Z^{-1}_{GGE} \exp{\left[\sum_k \frac{\omega_k}{\Delta}
       (\gamma_k^{\dagger} \gamma_k - \delta_k^{\dagger}
       \delta_k)\right]},
\ee
for $\Delta \gg 1$.  Although the above result may appear to be a thermal ensemble, 
we must recall the dispersion of the eigenmodes is not $\omega_k = -2 \cos k$ but 
 $E_k = \sqrt{\omega^2_k + \Delta^2}$. Thus, the temperature becomes again mode
 dependent and equal to $T(k) = \Delta E_k /\omega_k$, which  is consistent with the numerical results 
 of Ref.~\cite{Rigol07}, where lack of thermalization to the standard Gibbs ensemble but  thermalization to the GGE 
 was numerically found in this case. 

 Thus, it appears again that the emergence of thermal behavior in exactly solvable models requires, at least in the simplest case, 
 the existence of an energy gap in the spectrum of the Hamiltonian that determines the initial state.  
However, as we have already briefly mentioned in section~\ref{sec:example1}, the thermal ensemble
is just an approximation  to the more general GGE, which applies in all circumstances. 
Indeed, the GGE can reproduce the behavior of 
the asymptotic correlation functions  but it cannot reproduce the behavior of all observables~\cite{Cazalilla}.
This is because, in its simplest version of Eq.~\eqref{eq:GGE}, the GGE does not capture all the correlations 
between the eigenmodes that exist in the initial state. For example, in the example of section~\ref{sec:example1}, 
$\langle I_k I_{k+\pi} \rangle_{GGE} = \langle I_k \rangle_{GGE} \times \langle I_{k+\pi}\rangle_{GGE} 
\neq \langle\Phi_0 | I_k I_{k+\pi} | \Phi_0\rangle$, where $|\Phi_0\rangle$ is the initial 
state (i.e. the ground state of $H_0$, Eq.~\ref{eq:HRigolk}) and $I_{k+K} = f^{\dag}_{k+K} f_{k+K}$. 
This has important consequences, for
example, when  considering the energy fluctuations:
\begin{align}\label{eq:sigma}
\sigma^2 &=\langle \Phi_0 | H^2 | \Phi_0 \rangle - \langle\Phi_0 | H  | \Phi_0 \rangle^2 \\
&= \Delta^2\sum_k\frac{\omega^2_k}{E_k^2}. \label{eq:fluctt}
\end{align} 
On the other hand, if we compute the same quantity using the GGE, we find:
\begin{equation}
\sigma^2_{GGE} = \Delta^2\sum_k\frac{\omega^2_k}{2 E_k^2} = \frac{1}{2}\sigma^2. \label{eq:fluct}
\end{equation}
As we have shown in section~\ref{sec:example1},  for $\Delta \gg 1$, the GGE tends to a thermal Gibbs ensemble (TGE) with $T_{\mathrm{eff}} = \Delta/2= \beta^{-1}_{\mathrm{eff}}$  ($\rho_{GGE} \to \rho_{TGE}$). And  according to the fluctuation-dissipation theorem, for a thermal ensemble $\rho_{TGE}$
at a temperature $T_{\mathrm{eff}}$, 
 \begin{equation}
 C_V = \beta^{2}_{\mathrm{eff}}\: 
 \frac{\partial^2 \ln Z_{TGE}}{\partial \beta^2_{\rm eff}}\Bigg|_{V,\ldots} =  \frac{ \sigma^2_{\mathrm{TGE}}}{T^2_{\mathrm{eff}}},
 \end{equation}
where  $C_V$  the heat capacity of the system.  Yet,  the actual energy fluctuations of the system
following a quantum quench in which a superlattice of strength $\Delta\gg1$ is turned off at $t =0$  are given by $\sigma^2 = 2\sigma_{GGE} \simeq 2 \sigma_{TGE}$ (cf. Eq.~\ref{eq:fluct}). Therefore,  we conclude that the fluctuation-dissipation  theorem breaks down in the model of section~\ref{sec:example1}, in spite that the asymptotic correlations  appear to be essentially  thermal for $\Delta \gg 1$. 

 Nevertheless, although the effective temperature obtained in Eq.~\eqref{eq:efft} has no thermodynamic meaning in the  sense of the fluctuation-dissipation theorem,  it still represents a quantity that is worth determining experimentally. 
 The reason is that  $T_{\mathrm{eff}}$ is measure
 of the entanglement in the system. Indeed, determination of $T_{\rm eff}$ from, say, a measurement of the boson 
 momentum distribution, should allow for a determination of the entanglement spectrum of  $\mathcal{H}_{A(B)}$,
 which, according to the discussion in sections~\ref{sec:intro} and \ref{sec:example1}, is given by 
 $\mathcal{H}_{A(B)} \simeq H_{A(B)}/T_{\mathrm{eff}}$. Thus, an experimental determination of the 
 von Neumann  entropy, $S_{A(B)} = - \tr \rho_{A(B)} \log_2 \rho_{A(B)}$, ($\rho_{A(B)}  = e^{-\mathcal{H}_{AB}}/Z_{A(B)}$) 
 would be also possible. Similar remarks are applicable to the counter-example discussed in
 section~\ref{sec:example2}, provided we exchange the role of $T_{\mathrm{eff}}$ by $\epsilon$. 
 In this case, however, we cannot expect thermal correlations. 

In summary, we have presented a simple instance of  a quantum quench in which a quantum quench in
 an exactly solvable system  can produce  essentially thermal  correlations.  
The emergence of thermal correlations from the generalized Gibbs ensemble (GGE) has been
related to the existence of bi-partite eigenmode entanglement and 
a gap in the spectrum of the Hamiltonian that describes the initial
state. In this regard, we have also discussed a counter-example demonstrating that thermalization 
does not happen if the initial state is described by  a gapless Hamiltonian. Our results
allow to establish a link between the GGE and the entanglement spectrum in exactly solvable 
systems with bi-partite entanglement of the eigenmodes. Thus, it makes it possible an experimental 
measurement of the entanglement spectrum and other quantities derived from it (such like the 
von Neumann entropy), provided the asymptotic correlation functions of the system following 
a quantum quench can be measured  experimentally.  We have argued that this task 
becomes particularly simple when the GGE reduces to a thermal ensemble
(or, when the entanglement spectrum has a relatively simple form, as in the counter-example
discussed in section~\ref{sec:example2}). Finally, we have also shown that, even if correlations
may become essentially thermal, other quantities, such as the energy fluctuations, are not. 
This is akin to a  breakdown of the fluctuation-dissipation theorem.

\acknowledgments

  We are grateful to Prof. Ingo Peschel for a careful reading of the manuscript and many useful comments to improve the presentation.  MCC acknowledges the National Science Council of Taiwan. MAC thanks D.W. Wang for his hospitality at NCTS (Taiwan) and
A. H. Castro-Neto for his  hospitality at the Graphene Research Center at the National University of Singapore. MAC
acknowledges the financial support of the Spanish MEC through grant FIS2010-19609-C02-02.

\appendix

\section{Quantum quenches in the sine-Gordon model}

The sine-Gordon model has been introduced in the main text [see Eq. (\ref{eq:sG})]. This model has an exactly solvable point at $K=1$   (the so-called Luther-Emery point), at which  the Hamiltonian of Eq.~(\ref{eq:sG}) can be conveniently represented as a quadratic form of
 fermion fields. To this end, we must use the following bosonization identity:
\begin{equation}\label{eq:bosonization_formula}
\psi_\alpha(x)\sim\frac{1}{\sqrt{2\pi a}}\, e^{i s_\alpha\phi_\alpha(x)},
\end{equation}
where we have introduced the index $s_\alpha=1$ for $\alpha=R$ and $s_\alpha=-1$ for $\alpha=L$, and the chiral bosonic fields $\phi_\alpha = K^{-1/2}\phi+s_\alpha K^{1/2}\theta$, and $\psi_{R,L}$ are destruction operators for spinless fermions moving to the right and to the left, respectively. Using the above identity and, after a Fourier transformation, the Hamiltonian becomes:
\begin{multline}
H_{sG}  =\sum_{p}\varepsilon_{p}^{0}\left[\psi_{R}^{\dagger}(p)\psi_{R}(p)-\psi_{L}^{\dagger}(p)\psi_{L}(p)\right] \\
+\Delta\sum_{p}\left[\psi_{R}^{\dagger}(p)\psi_{L}(p)+\psi_{L}^{\dagger}(p)\psi_{R}(p)\right],\label{eq:sG_fermions2}
\end{multline}
with linear dispersion $\varepsilon_{p}^{0}= v p$.

Let us consider a quench in which the system is initially prepared in
the gapped ground state of $H_\mathrm{sG}$, or more generally, in a thermal
state defined by a density operator $\rho=e^{- H_\mathrm{sG} /T}$ with  temperature $T$. We then assume that the coupling $g$ is suddenly turned off at $t=0$, and therefore, for $t>0$ the time evolution is governed by $H_0$. In the long times regime, the expectation value and correlations of a broad class of operators for long times can be described by the GGE~\cite{Iucci2}.

The sine-Gordon Hamiltonian at the Luther-Emery point can be diagonalized by the Bogoliubov transformation
\begin{align}
\psi_{R}(p) & =\cos\theta_{p}\psi_{c}(p)-\sin\theta_{p}\psi_{v}(p)\\
\psi_{L}(p) & =\sin\theta_{p}\psi_{c}(p)+\cos\theta_{p}\psi_{v}(p)
\end{align}
provided we choose
\begin{equation}
\tan2\theta_{p}  =\frac{\Delta}{\varepsilon_{p}^{0}}.
\end{equation}
In terms of the new variables $\psi_{v,c}$ it turns out to take the diagonal form
\begin{equation}
H_{sG}=\sum_{p}\varepsilon_{p}\left[\psi_{c}^{\dagger}(p)\psi_{c}(p)-\psi_{v}^{\dagger}(p)\psi_{v}(p)\right].
\end{equation}
with dispersion $\varepsilon_{p}  =\sqrt{[\varepsilon_{p}^{0}]^{2}+\Delta^{2}}$. 
This Hamiltonian is a continuum version of the superlattice model discussed in section~\ref{sec:example1},
which is obtained from \eqref{eq:HRigolk} in the limit where $k=p \to 0$.

For an initial  thermal state, the eigenmode occupations are: 
\begin{align}
\langle\psi_{R}^{\dagger}(p)\psi_{R}(p)\rangle &
=\frac{1}{2}\left(1-\cos2\theta_{p}\tanh\frac{\varepsilon_{p}}{2 T}\right)\\
\langle\psi_{L}^{\dagger}(p)\psi_{L}(p)\rangle &
=\frac{1}{2}\left(1+\cos2\theta_{p}\tanh\frac{\varepsilon_{p}}{2 T}\right)
\end{align}
and hence the values of $\alpha$ and $\beta$ that determine the GGE read:
\begin{align}
\alpha(p)&
=\log\left[\frac{1-\cos2\theta_p\tanh\frac{\Delta}{2 T}}{1+\cos2\theta_p\tanh\frac{\Delta}{2
    T}}\right]\\
\beta(p) &
=\log\left[\frac{1+\cos2\theta_p\tanh\frac{\Delta}{2 T}}{1-\cos2\theta_p\tanh\frac{\Delta}{2T}}\right].
\end{align}



\begin{thebibliography}{30}
\bibitem{Weiss}
T. Kinoshita, T. Wenger, and D. S. Weiss, Nature (London)
440, 900 (2006).
%
\bibitem{Schmiedmayer}
S. Hofferberth \emph{et al.} Nature (London) {\bf 449}, {\bf 324} (2007).
%
\bibitem{Bloch}
M. Greiner \emph{et al.}, Nature (London) {\bf 419}, 51 (2002).
S. Trotsky \emph{et al.} arxiv: 1101.2658 (2011).
%
\bibitem{Greiner}
J Simon, W S Bakr, R Ma, M E Tai,  P M Preiss, M Greiner,
{\bf 472},  307, Nature (2011)
%
\bibitem{CazalillaRigol}
M. A. Cazalilla and M. Rigol, New J. of Phys. {\bf 12} 055006 (2010).
%
\bibitem{Rigol07}
M. Rigol, V. Dunjko, V. Yurovsky, and M. Olshanii, Phys. Rev.
Lett. {\bf 98}, 050405 (2007).
\bibitem{Rigol06} M. Rigol,A. Muramatsu and M. Olanshii, Phys. Rev. A
  {\bf 74}, 053616 (2006)
\bibitem{Rigol11}
M. Rigol and M. Fitzpatrick, Phys. Rev. A {\bf 84}, 033640 (2011).
%
\bibitem{Sengupta}
K. Sengupta, S. Powell, and S. Sachdev, Phys. Rev. A {\bf 69}, 053616 (2004).
%
\bibitem{Cazalilla}
M. A. Cazalilla, Phys. Rev. Lett {\bf 97} 156403 (2006); A. Iucci and M.~A. Cazalilla, Phys. Rev. A {\bf 80}, 063619 (2009).
%
\bibitem{attention}
E. Altman and A. Auerbach, Phys. Rev. Lett. {\bf 89}, 250404 (2002);
 P. Calabrese and J. Cardy, \emph{ibid} {\bf 96}, 136801 (2006);
S.~R. Manmana, S. Wessel, R. M. Noack, and A. Muramatsu,
\emph{ibid} {\bf 98} 210405 (2007);
M. Eckstein and M. Kollar,  \emph{ibid}  {\bf 100}, 120404
(2008); P. Reimann,  \emph{ibid}  {\bf 101}, 190403 (2008);
P. Barmettler, M. Punk, V. Gritsev, E. Demler, and E. Altman,
 \emph{ibid} {\bf  102}, 130603 (2009);  M.  Moeckel and
 S. Kehrein  \emph{ibid} {\bf 100} 175702 (2008);
 D. Sen, K. Sengupta, and
 S. Mondal,  \emph{ibid}  {\bf 101}, 016806 (2008);
 A. Faribault, P. Calabrese, and J.-S. Caux, J. Stat. Mech.:
Theory Exp. (2009) P03018; C. De Grandi, V. Gritsev, A. Polkovnikov
 Phys. Rev. B {\bf 81}, 012303 (2010);  \emph{ibid} {\bf 81}, 224301 (2010);
 L. F. Santos, M. Rigol, and  A. Polkovnikov, arXiv:1103.0557 (2011).
%
\bibitem{MiguelAnibalMing}
M.A. Cazalilla, A. Iucci and M.-C. Chung,
Phys. Rev. E {\bf 85}, 011133 (2012).
%
\bibitem{Polkovnikovetal}
See A. Polkovnikov, K. Sengupta, A. Silva, M. Vengalattore, Rev. Mod. Phys. \textbf{83}, 863 (2011) for a review.
%
\bibitem{grisins} P. Grisins, I. E. Mazets, Phys. Rev. A \textbf{84}, 053635 (2011).
%
\bibitem{afho10}
S. Genway, A.~F. Ho, and D.~K.~K. Lee, Phys. Rev. Lett. {\bf 260402} (2010)
%
\bibitem{Rigol_ETH}
M. Rigol, V. Dunjko, and M. Olshanii, Nature (London) {\bf 452}, 854 (2008).
%
\bibitem{gring} M Gring \emph{et al.}, arXiv:1112.0013
%
\bibitem{mitraGiam} A. Mitra and T. Giamarchi, Phys. Rev. Lett. \textbf{107}, 150602 (2011).
%
\bibitem{Peschel} for a review, see: I. Peschel and V. Eisler,
  J. Phys. A : Math. Theor., {\bf 42}, 504003 (2009)
%
\bibitem{Li/Haldane08} H. Li and F.D.M. Haldane,
 Phys. Rev. Lett., {\bf 101}, 010504 (2008)
%
\bibitem{ChungJhuChenYip} M.-C. Chung, J.-H Jhu, P. Chen and S.K. Yip,
Eur. Phys. Lett. {\bf 95}, 27003 (2011)
%
\bibitem{QiLudwig} X.-L. Qi, H. Katsura and A.W.W. Ludwig, arXiv:1103.5437
%
\bibitem{Pollmann} F. Pollmann, E. Berg, A.M. Turner and M. Oshikawa,
  Phys. Rev. B {\bf 81}, 064439 (2010)
%
\bibitem{Lauchli10} A.M. L\"auchli., E.J.  Bergholtz, J. Suorsa  and M. Haque, Phys. Rev. Lett., {\bf 104} 156404 (2010).
%
\bibitem{Schliemann11} J. Schliemann, Phys. Rev. B, {\bf{83}} 115322 (2011).
%
\bibitem{Poilblanc10} D. Poilblanc, Phys. Rev. Lett. , {\bf 105}, 077202 (2010).
%
\bibitem{PeschelChung11} I. Peschel and M.C. Chung, Eur. Phys. Lett.  {\bf 96}, 50006 (2011).
%
\bibitem{ReducedDensity} M. C. Chung and I. Peschel, Phys. Rev. B {\bf
    64}, 064412 (2001); I. Peschel, J. Phys. A {\bf 36}, L205 (2003); S. A. Cheong and C. L. Henley, Phys. Rev. B 69, 075111 (2004); T. Barthel {\it et. al} Phe. Rev. A {\bf 74}, 022329 (2006).
%
\bibitem{giamBook} T. Gimarchi, \emph{Quantum Physics in One Dimension} (Oxford University Press, Oxford, 2004).
%
\bibitem{lieb_antiferromagnetic_chain} E. Lieb, T. Shulz and D. Mattis, Ann. Phys. (N. Y.), \textbf{16}, 407 (1961).
%
\bibitem{Iucci2} A. Iucci and M.~A. Cazalilla, New Journal of Physics {\bf 12}, 055019 (2010).

%









\end{thebibliography}
\end{document}